# Ptycho-endoscopy on a lensless ultrathin fiber bundle tip

Pengming Song[1,†,*], Ruihai Wang[1,†], Lars Loetgering[3], Jia Liu[1], Peter Vouras[4], Yujin Lee[5], Shaowei Jiang[1], Bin Feng[1], Andrew Maiden[6,7], Changhuei Yang[8], and Guoan Zheng[1,2*]

[1]Department of Biomedical Engineering, University of Connecticut, Storrs, CT 06269, USA
[2]Center for Biomedical and Bioengineering Innovation, University of Connecticut, Storrs, CT 06269, USA
[3]CarlZeiss AG, Carl Zeiss Promenade, Jena, Thuringia 07745, Germany
[4]United States Department of Defense, Washington, D.C. 20301, USA
[5]School of Electrical and Electronic Engineering, Yonsei University, Seoul 03722, Republic of Korea
[6]Department of Electronic and Electrical Engineering, University of Sheffield, Sheffield, South Yorkshire S1 3JD, UK
[7]Diamond Light Source, Harwell, Oxfordshire OX11 0DE, UK
[8]Department of Electrical Engineering, California Institute of Technology, Pasadena, CA 91125, USA
[†]These authors contributed equally to this work
[*]Correspondence: pengming.song@uconn.edu (P.S.) or guoan.zheng@uconn.edu (G.Z.)

**Abstract:** Synthetic aperture radar (SAR) utilizes an aircraft-carried antenna to emit electromagnetic pulses and detect the returning echoes. As the aircraft travels across a designated area, it synthesizes a large virtual aperture to improve image resolution. Inspired by SAR, we introduce synthetic aperture ptycho-endoscopy (SAPE) for micro-endoscopic imaging beyond the diffraction limit. SAPE operates by hand-holding a lensless fiber bundle tip to record coherent diffraction patterns from specimens. The fiber cores at the distal tip modulate the diffracted wavefield within a confined area, emulating the role of the 'airborne antenna' in SAR. The handheld operation introduces positional shifts to the tip, analogous to the aircraft's movement. These shifts facilitate the acquisition of a ptychogram and synthesize a large virtual aperture extending beyond the bundle's physical limit. We mitigate the influences of hand motion and fiber bending through a low-rank spatiotemporal decomposition of the bundle's modulation profile. Our tests demonstrate the ability to resolve a 548-nm linewidth on a resolution target. The achieved space-bandwidth product is ~1.1 million effective pixels, representing a 36-fold increase compared to that of the original fiber bundle. Furthermore, SAPE's refocusing capability enables imaging over an extended depth of field exceeding 2 cm. The aperture synthesizing process in SAPE surpasses the diffraction limit set by the probe's maximum collection angle, opening new opportunities for both fiber-based and distal-chip endoscopy in applications such as medical diagnostics and industrial inspection.

# Introduction

Synthetic aperture radar (SAR) has revolutionized the field of remote sensing by generating detailed terrestrial images from airborne platforms[1]. Its profound impact spans across environmental monitoring, disaster management, military surveillance, and planetary exploration, where SAR consistently delivers unparalleled high-resolution images of landscapes. In a typical SAR system, an aircraft-carried antenna emits a series of microwave pulses to illuminate a target scene. The same antenna then records the returning echoes to infer the distance to the target. As the aircraft moves along its flight path, the successive echoes acquired at multiple positions can be processed to form a large virtual aperture. This synthesized aperture significantly exceeds the physical dimension of the antenna, allowing the creation of higher-resolution images than would otherwise be possible. For data acquisition, SAR utilizes a common phase reference to ensure coherent detection of echoes acquired at different positions along the flight path. This reference is often provided by a highly stable local oscillator onboard the aircraft. Despite its success, implementing the concept of SAR at optical wavelengths is challenging due to the absence of a suitable local oscillator and the stringent opto-mechanical requirements for interferometric measurements. Notably, efforts have been made to apply the SAR concept in the optical domain, such as in the Laser Interferometer Space Antenna (LISA) mission[2], which aims to detect gravitational waves using laser interferometry.

In parallel to the development of SAR, ptychography has emerged as a pivotal lensless diffraction imaging technique for both fundamental and applied sciences[3, 4, 5, 6, 7, 8, 9, 10]. Unlike SAR, ptychography operates at shorter wavelengths from X-ray to visible light, where direct temporal measurement of coherent wavefields is often impossible. In a typical ptychographic implementation, a spatially confined probe beam illuminates the object in real space, and the intensity of the resulting diffracted light waves is acquired using an image sensor in reciprocal space. By translating the object to different lateral positions, ptychography acquires a set of diffraction patterns collectively referred to as a ptychogram[5]. Each recorded pattern in reciprocal space contains object information tied to the corresponding spatially confined region in real space. The subsequent reconstruction process stitches these confined regions together to expand the imaging field of view, and simultaneously recovers both the amplitude and phase of the specimen's complex wavefield. A key aspect of ptychography is the intentional overlap of the illumination areas during the object translation process. This overlap ensures the object interacts multiple times with the illumination probe beam, thereby resolving phase ambiguities and expediting the recovery process[4].

Given that ptychography obviates the need for interferometric measurements and allows for imaging extended objects without compact support constraints, it has rapidly attracted interest from different scientific communities. In the realm of coherent X-ray microscopy, ptychography has become an indispensable imaging tool in most synchrotron and national laboratories worldwide[11]; in electron microscopy, recent advances have pushed the resolution to record-breaking deep sub-angstrom levels[12]; and in optical imaging, spatial- and Fourier-domain ptychography has provided turnkey solutions for high-resolution, high-throughput microscopy with minimal hardware modifications[13, 14].

Drawing on the principles of SAR and ptychography, we present Synthetic Aperture Ptycho-Endoscopy (SAPE), a non-interferometric approach that achieves super-resolution imaging beyond the diffraction limit of the lensless endoscopic probe. SAPE uses a handheld lensless fiber bundle tip to record coherent diffraction patterns from a specimen. The fiber bundle tip serves two roles in SAPE. First, it relays the object's diffracted wavefield from the distal end to the proximal end for intensity-based detection, mimicking the role of the aircraft-carried antenna in SAR. Second, the densely-packed fiber cores at the distal tip modulate both the amplitude and phase of the diffracted wavefield within a confined area of the bundle facet, akin to the probe beam's spatially-confined modulation in ptychography. The handheld operation also serves two roles in SAPE. First, it introduces positional shifts to the fiber bundle tip for synthesizing a large virtual aperture, analogous to the aircraft's movement in SAR. Second, the shifts of the distal tip introduce transverse translation diversity[15] for the acquisition of a ptychogram, analogous to the object translation process in ptychography. Notably, the SAPE concept can also be extended to distal-chip endoscopy, where a miniaturized camera is directly placed at the distal end of the endoscope probe[16, 17]. To implement SAPE in this setting, the small lens element of the miniaturized camera can be replaced by a thin coded layer on the sensor[18], which modulates the diffracted wavefields like the fiber bundle tip.

In SAPE, the achievable resolution is not constrained by the maximum collection angle of the distal imaging probe, a limitation encountered in current endoscopic techniques. Instead, resolution is primarily determined by the translation distance of the handheld endoscope probe. Unlike SAR, which relies on the direct coherent acquisition of a complex wavefield through a local oscillator, SAPE simplifies this process by only capturing the intensity patterns of the object's diffracted wavefields. The phase information is recovered via ptychographic reconstruction post-measurement. This eliminates the need for a local oscillator, which is particularly advantageous when operating at optical wavelengths where such oscillators are not readily available. Furthermore, SAPE distinguishes itself from interferometric or holographic approaches[19, 20, 21, 22, 23, 24, 25, 26, 27] by eliminating the need for a reference wave. This streamlines the experimental setup and boosts the system's adaptability across diverse settings. The digital refocusing feature of SAPE further facilitates the reconstruction of the specimen's three-dimensional complex topographic profile, a functionality akin to the measurement of terrestrial heights in interferometric SAR[28]. Different from other endoscopic imaging methods[29, 30, 31, 32, 33, 34, 35, 36, 37], SAPE heralds new opportunities for synthetic aperture endoscopic imaging beyond traditional optical constraints. It also unlocks new possibilities for implementing the SAR concept at optical wavelengths, expanding the scope for remote sensing with autonomous vehicles and drones.

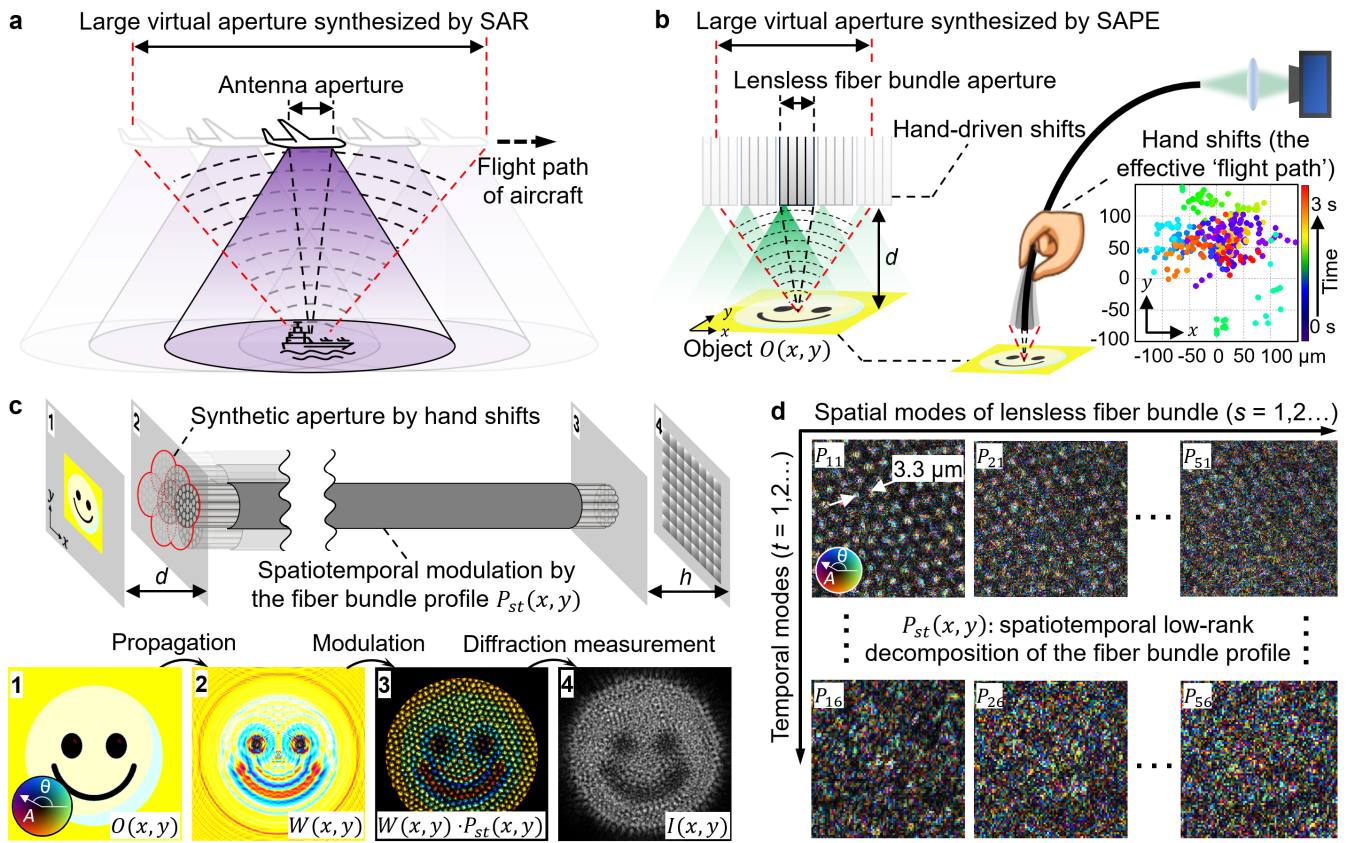

**Fig. 1| Principle of SAPE.** (a) In SAR, an airborne antenna emits a series of electromagnetic pulses to illuminate a target scene, and the returning echoes are processed to synthesize a large virtual aperture that exceeds the physical size of the antenna. (b) SAPE uses a handheld fiber bundle tip as a 'moving antenna' to illuminate the object and record the resulting diffraction patterns. The natural hand motion induces shifts at the fiber bundle tip, mirroring the aircraft's movement in SAR. These shifts enable ptychographic data collection and the creation of an expanded synthetic imaging aperture extending beyond the bundle's physical limit. (c) Imaging model of SAPE. Panel 1: The object is illuminated with an extended beam in either a transmission or reflection setting. Panel 2: The object's diffracted wavefield reaches the plane of the distal fiber bundle tip. Alternatively, when implementing SAPE for distal-chip endoscopy, a thin coded layer can replace the small lens of the miniaturized camera at the distal end, serving the same purpose of modulating the diffracted wavefield. Panel 3: The fiber bundle tip or coded surface modulates the diffracted wavefield over a confined region, acting as an equivalent spatially-confined probe beam in ptychography. Panel 4: The modulated wavefield, upon exiting the proximal end of the bundle, travels a distance $h$ before its intensity is recorded by an image sensor. For distal-chip endoscopy, this distance corresponds to the space between the coded surface and the pixel array of the image sensor. (d) SAPE compensates for hand motions and fiber bending by decomposing the modulation profile into low-rank spatiotemporal modes $P_{st}(x, y)$.

## Results

### Principle of SAPE

Figure 1a shows the operation of SAR, where an airborne antenna emits a series of electromagnetic pulses and subsequently detects the returning echoes from the target scene. Signal processing of the successive echoes forms a large virtual aperture that exceeds the physical dimension of the antenna. Inspired by this principle, Figure 1b shows the schematic of SAPE, where a handheld lensless fiber bundle acts as a 'moving antenna' to acquire coded diffraction patterns from the object $O(x, y)$. The handheld operation inherently introduces translational shifts ($x_j$, $y_j$) to the distal tip, facilitating the acquisition of a ptychogram. Notably, these shifts can be recovered post-measurement through a multi-step correlation analysis process of the captured diffraction patterns[13, 38], eliminating the need for precise real-time positional tracking as in conventional ptychography (Supplementary Note 1 and Figs. S1-S2). The bottom right panel of Fig. 1b illustrates the positional shifts of the fiber bundle tip recovered from a SAPE experiment. The data points are color-coded to indicate their respective time stamps.

Figure 1c illustrates the imaging model of SAPE, with panels 1-4 visualizing the wavefields at successive stages, where the amplitude and phase are respectively represented by grey levels and color hues. Unlike the spatially confined probe beam used in conventional ptychography, SAPE illuminates the object $O(x, y)$ with an extended beam, which can be a collimated plane wave or a spherical wave emanating from a single-mode fiber core. The extended beam interacts with the object and the resulting wavefield propagates to the plane of the distal fiber bundle tip. This diffracted wavefield, represented as $W(x, y)$, is shown in panel 2 of Fig. 1c. The densely packed fiber cores at the tip then modulate both the amplitude and phase of $W(x, y)$, similar to the object-probe modulation process in conventional ptychography. The modulated wavefield subsequently exits the proximal end of the fiber bundle and further propagates a distance $h$ before being detected by an image sensor. Light diffraction by a small distance is necessary as it can convert the phase information into intensity variations for detection[39] (Supplementary Note 2 and Fig. S3).

With Fig. 1c, the forward imaging model for SAPE can be expressed as:

$$I_j(x, y) = \sum_s \left| \{W(x, y) \cdot FB_{sj}(x - x_j, y - y_j)\} * psf_{free}(h) \right|^2 \quad (1)$$

Here, $I_j(x, y)$ denotes the $j^{th}$ acquired diffraction pattern corresponding to the distal fiber bundle shift ($x_j$, $y_j$). $W(x, y)$ represents the object's diffracted wavefield at the plane of the lensless distal fiber bundle tip. $FB_{sj}(x, y)$ represents the fiber bundle's modulation profile, where the subscript 's' denotes different spatially incoherent modes induced by hand motion, and 'j' specifies the mode profile at the $j^{th}$ measurement timestamp. The convolution kernel $psf_{free}(h)$ models free-space light propagation over a distance $h$. It captures the diffraction effects that occur after the wavefield exits the proximal end of the fiber bundle, as illustrated in Fig. 1c. The symbol '·' indicates element-wise multiplication, and '∗' denotes convolution. In Eq. (1), the captured image is essentially a composite of spatially incoherent modes induced by motion, as indicated by the subscript 's' next to the summation symbol.

Directly recovering the position-dependent modulation profile $FB_{sj}(x, y)$ is a challenging task due to the extensive degrees of freedom involved. In SAPE, we simplify this problem by assuming that the variations of the fiber bundle modulation profile are gradual and minor over time, which allows for a low-rank spatiotemporal decomposition of the profile[10, 40]. Specifically, we reconstruct a series of fiber bundle profiles $FB_{sj}(x, y)$ for $J$ different shift positions and project these $J$ profiles into a lower $T$-dimensional space via singular value decomposition (SVD), where $T \ll J$:

$$[P_{s[.]}, D_s, V_s] = truncated\_SVD\{FB_{s[.]}, T\} \quad (2)$$

In Eq. (2), $FB_{s[.]}$ is formed by converting each 2D profile of $FB_{sj}(x, y)$ into a vector and then concatenating all $J$ vectors column-wise. $P_{s[.]}$ is the projected orthogonal matrix with $T$ columns, and each column can be reshaped into a fiber bundle orthogonal mode $P_{st}(x, y)$, with $t = 1, 2, \ldots T$. $D_s$ is a $T$-by-$T$ diagonal matrix of singular values, and $V_s$ is a $J$-by-$T$ complex conjugated orthonormal evolution matrix. With the orthogonal mode $P_{st}(x, y)$, the $j^{th}$ captured image can be rewritten as:

$$I_j(x, y) = \sum_s \left| \{W(x, y) \cdot \sum_t (P_{st}(x - x_j, y - y_j) \cdot \alpha_{tj})\} * psf_h(x, y) \right|^2 \quad (3)$$

Here, $\alpha_{tj} = (D_s \times V_s^H)_{tj}$, presenting the element in the $t^{th}$ row and the $j^{th}$ column of the matrix product between $D_s$ and the Hermitian transpose of $V_s$. In distal-chip endoscopy, a miniaturized camera is attached to the distal end for direct image acquisition. As Eq. (3) suggests, an alternative implementation of SAPE is to replace the small lens of the miniaturized camera with a coded surface directly attached on top of the coverglass of the image sensor[18]. The coded surface serves the role of $P_{st}$ in Eq. (3), modulating the diffracted wavefield from the object. The term '$* psf_h(x, y)$' in Eq. (3) then models light propagation from the coded surface to the pixel array of the image sensor. With the captured images $I_j(x, y)$ ($j = 1, 2, ...J$), SAPE aims to jointly recover the high-resolution object diffracted wavefield $W(x, y)$, and the spatiotemporal orthogonal modes $P_{st}(x, y)$. A detailed discussion of the recovery process can be found in Supplementary Note 3.

A key distinction between SAPE and conventional ptychography resides in the methodology of information recovery, as elaborated in Supplementary Note 4 and Fig. S4. Conventional ptychography focuses on reconstructing object information in real space by effectively stitching together regions illuminated by the confined probe beam, thereby expanding the imaging field of view. In contrast, SAPE aims to recover the diffracted wavefield $W(x, y)$ at the plane of the distal endoscope facet, essentially stitching information in a diffraction plane as in synthetic aperture ptychography[41]. This process is illustrated in Figs. 1b and 1c, where the synthetic aperture surpasses the physical limit of the endoscope probe. Different from conventional ptychography with a confined probe beam for illumination, the lateral translation shift of the imaging probe in SAPE is very small compared to the dimension of the probe. Therefore, the overlap rate of SAPE often ranges from 95% to 99%, similar to that in coded ptychography[42, 43, 44]. Furthermore, conventional spatial-domain and Fourier-domain ptychography often necessitate a thin sample assumption[3, 6, 9, 45]. This assumption allows the interaction between the object and the structured probe beam to be approximated through point-wise multiplication[4]. In contrast, SAPE eliminates the need to model the interaction between the object and the illumination beam. The diffracted wavefield $W(x, y)$ in SAPE depends solely on how the complex-valued light waves exit the object, not entering it[13, 41, 46, 47]. This allows for digital propagation of the diffracted wavefield to various axial positions[44, 47], thereby facilitating the imaging of objects with arbitrary thickness or topographic profiles. By maximizing the image contrast or other metrics during digital refocusing[48], we can further recover the 3D topographic profiles of the samples.

**Experimental validation and characterization**
To validate the performance of SAPE, we conducted imaging tests on a diverse set of samples. Figure 2 illustrates the experiments performed on a resolution target and a blood smear slide in a transmission configuration, where the specimens were illuminated with an extended plane wave. Figure 2a shows the experimental setup, wherein a handheld ultrathin fiber bundle is positioned approximately ~50 microns from the specimen to acquire the diffraction patterns. The fiber bundle has a diameter of ~650 µm and a core-to-core distance of ~3.3 µm. As shown in Figs. 2b and 2c, the captured raw images of both the resolution target and the blood smear slide exhibit noticeable pixelation. This issue arises from the discrete nature of the fiber cores and the relatively large spacing between them. It can constrain the resolution and obscure finer details within the specimen. Figure 2d reveals the SAPE reconstruction of the resolution target, where it effectively addresses the pixelation effect and resolves a linewidth of 870 nm in group 9, element 2. The image acquisition process is featured in Supplementary Video S1. In Fig. 2e, we demonstrate the SAPE reconstruction of the blood smear slide, revealing both the amplitude and phase properties of individual blood cells. For the reconstructions in Figs. 2d and 2e, we performed a low-rank decomposition of the modulation profile to mitigate the issues of hand motion and fiber bending. Supplementary Video S2 illustrates the dynamic evolution of the fiber bundle's modulation profile throughout the acquisition process. The single-mode nature of the fiber cores ensures minimal variation in the modulation profile over time.

In an additional experiment, we translated the distal fiber bundle tip with a motorized stage for data acquisition, recording the stage positions at each step. As shown in Supplementary Fig. S5, the resulting resolution has been improved to resolve group 9, element 6 on the resolution target, corresponding to a linewidth of 548 nm. The distance between the fiber bundle tip and the resolution target is ~50 microns. This represents one of the highest resolutions attained using a fiber bundle, marking a 6-fold resolution gain compared to the 3.3-µm core-to-core distance of the lensless fiber bundle tip. The space-bandwidth product of the employed fiber bundle is 30,000 (with 30,000 fiber cores). If we don't consider the expansion of the field of view, SAPE achieves ~36-fold increase in the space-bandwidth product, resulting in a recovered image with ~1.1 million effective pixels within the same area of the fiber bundle.

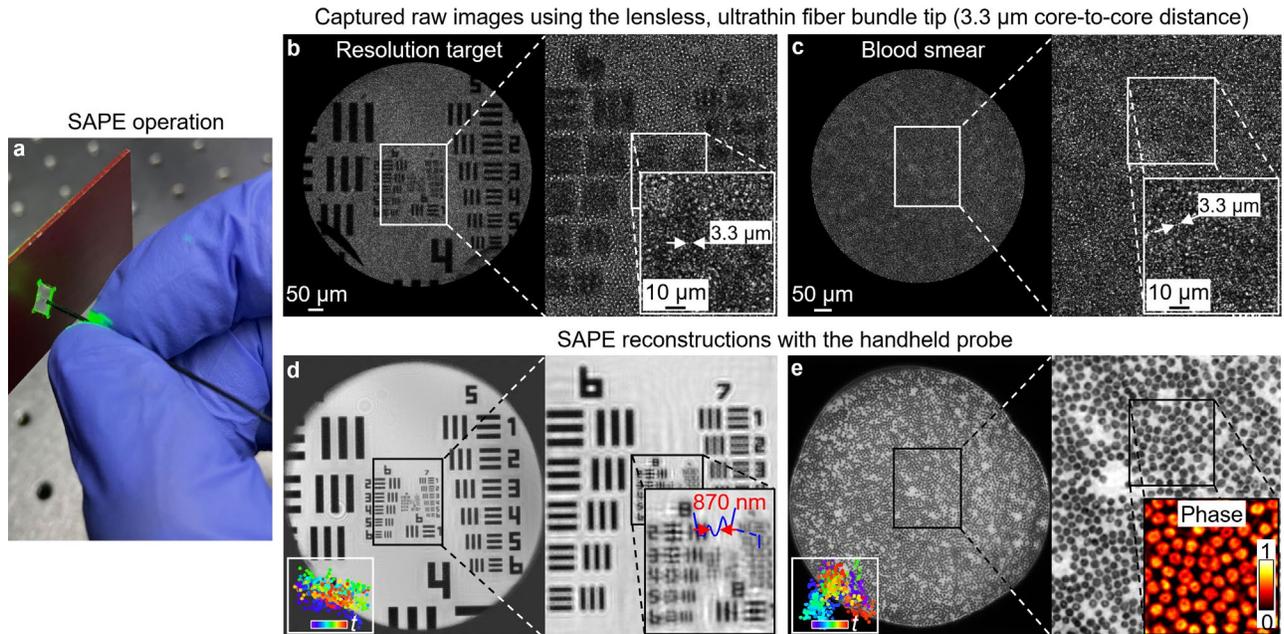

**Fig. 2| Validation and characterization of SAPE in a transmission configuration.** (a) SAPE employs a handheld, lensless fiber bundle to acquire diffraction patterns from the object. (b) The captured raw image of a resolution target using the fiber bundle with a 3.3-μm core-to-core distance. (c) The captured raw image of a blood smear slide. (d) SAPE reconstruction of the resolution target, successfully resolving the 870-nm linewidth on the target (the width of the entire group 9, element 2 is 4.4 μm). By translating the fiber bundle with a motorized stage, SAPE can further resolve the 548-nm linewidth in Supplementary Fig. S5. (e) SAPE reconstruction of both amplitude and phase properties of the blood smear slide. The bottom left insets in (d) and (e) reveal the recovered trajectory of the handheld fiber bundle tip post-measurement. Supplementary Video S1 shows the acquisition process of SAPE and Video S2 shows the dynamic modulation profile of the fiber bundle during acquisition.

In Supplementary Fig. S6, we also tested a hair-thin lensless fiber bundle with a diameter of 200 μm and the same object-to-bundle distance. The obtained results demonstrate the ability to resolve up to group 9, element 5 on the resolution target, corresponding to a linewidth of 620 nm, despite the hair-thin size of the imaging probe. Furthermore, Supplementary Fig. S7 provides a comparative analysis of SAPE reconstructions with and without performing spatiotemporal decomposition on the modulation profile. This demonstration reveals a notable degradation in image quality when a consistent modulation profile is applied across all translated positions.

Further validation of SAPE was conducted in a reflection configuration in Fig. 3a. In the reflection setup of SAPE, a separate single-mode fiber or a collimated beam can be used to deliver laser light for object illumination. The reflected light from the specimens can then be collected by the fiber bundle tip. Figures 3b-3d show the captured raw images of a resolution test target, a micro-circuit chip, and an ex-vivo mouse colon sample. The SAPE reconstructions are presented in Figs. 3e-3g, where amplitude information is represented by the grey levels while the phase is represented by color hues. These recovered images enable us to resolve the 1.4-μm linewidth from the resolution target, identify defects on the micro-circuit chip, and visualize the colonic crypts on the mouse colon. In contrast to the transmission configuration in Fig. 2, the specimens in Fig. 3 were positioned at 1.5-3 mm from the fiber bundle tip so that the illumination beam covered a larger area. If the specimen is placed too close to the fiber bundle tip, the tip could obstruct the illumination light. In Supplementary Fig. S8, we show the entire captured image of the resolution target, which highlights the shadow region caused by the fiber bundle's blockage. An alternative approach for illumination is to couple light into one of the fiber cores within the fiber bundle itself. However, this configuration may require additional optical components, such as a beam splitter and polarization optics, to minimize direct light reflection from the proximal end of the fiber bundle. Another consideration of this approach is that the specimen may need to be positioned at a greater distance from the fiber bundle tip to ensure adequate coverage of the illumination beam. For the implementation of SAPE in distal-chip endoscopy, where a miniaturized coded sensor is directly integrated at the distal end of the endoscope, we can incorporate a small beam splitter on top of the coded sensor chip. This beam splitter would be responsible for directing the collimated illumination beam onto the sample. Supplementary Fig. S9 presents a comparison between images obtained using SAPE and those captured with a conventional lens-based microscope equipped with a 10×, 0.4 NA objective lens. Both techniques effectively identify defects in micro-circuits. For the experiments in Figs. 2-3, the acquisition time was 3 seconds, involving 300 raw images captured at 100 frames per second. In Supplementary Figs. S10-S11, we further explore SAPE reconstructions with varying numbers of raw measurements, ranging from 50 to 500, corresponding to acquisition times between 0.5 to 5 seconds.

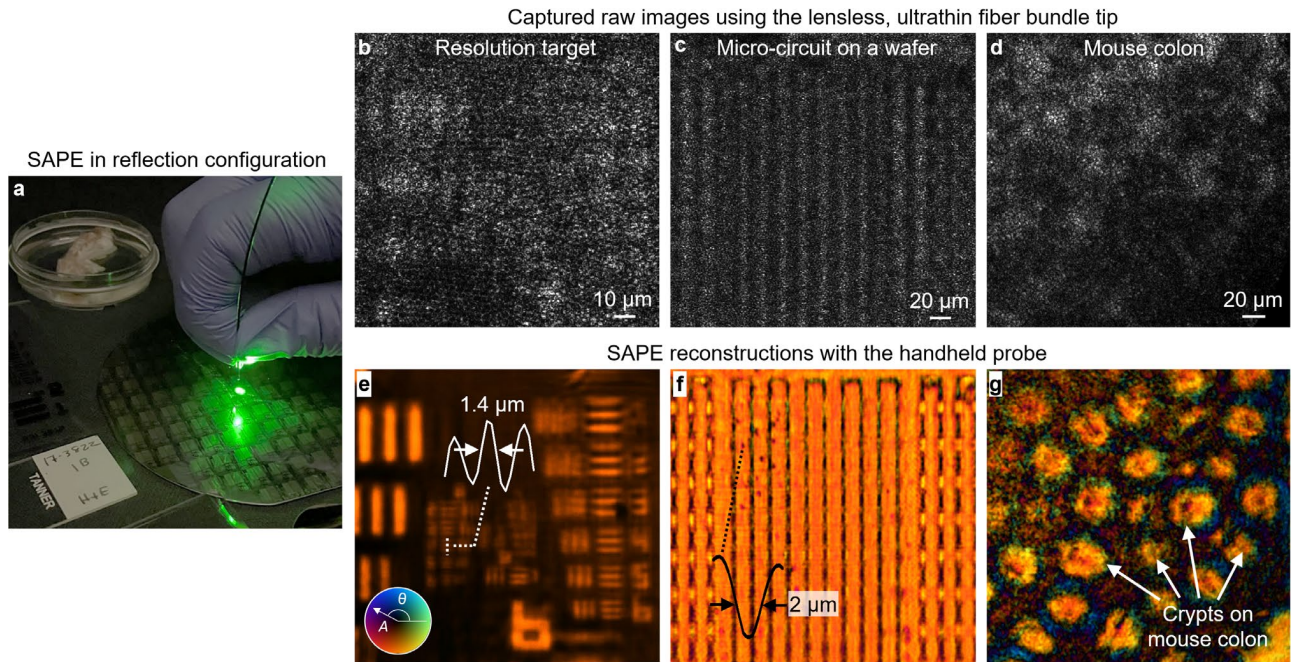

**Fig. 3| SAPE imaging in a reflection mode.** (a) In the reflection setup of SAPE, an external collimated beam or light from a single-mode fiber core can be used for object illumination. (b)-(d) The captured raw images of a resolution target, a micro-circuit chip, and a mouse colon sample. (e)-(g) SAPE reconstructions with amplitude represented by grey values while phase by color hues. The width of the entire group 8, element 4 in (e) is 7 μm.

**3D topographic imaging by SAPE**
Another distinct benefit of SAPE is post-measurement digital wavefield propagation, which facilitates the imaging of specimens with complex topographical variations – a task that poses difficulties for traditional imaging techniques. In Fig. 4a-4c, we used a tilted blood smear slide to demonstrate SAPE's capability of 3D digital refocusing in transmission mode. Upon recovering the object's complex wavefield at the plane of the distal end, we digitally propagate it to different axial distances for post-measurement refocusing. Figure 4a shows the ptychographic reconstruction refocused at 210 μm towards the object, where the top right portion is in focus. Figure 4b presents an all-in-focus image constructed by stitching together the in-focus portions of the digital refocused wavefields, where the depth information is encoded in the color hue. This synthesized view eliminates the need for manual refocusing or mechanical adjustments during the image acquisition process, addressing a common challenge in conventional lens-based systems. Figure 4c shows the zoomed-in views of Figs. 4a-4b, providing a closer look at the details of the digital refocusing process. Accompanying this figure, Supplementary Video S3 shows the digital propagation process at various axial distances, underlining the versatility of SAPE in imaging specimens with non-planar topographic profiles.

For industrial inspection applications, it is important to detect and visualize small-scale defects, irregularities, and surface variations in manufactured parts and components. Conventional industrial endoscopes with distal cameras have limited resolution and cannot recover the 3D topographic profile of the surface. The SAPE concept can be adopted in distal-chip endoscopes to enable high-resolution industrial inspection. In Fig. 4d-4f, we demonstrate the potential of SAPE for lensless 3D topographic imaging. In this demonstration, we first recovered the complex wavefield at the distal end of the lensless endoscope probe and then propagated this wavefield toward the object for digital refocusing. The resulting ptychographic reconstruction at the object plane is shown in Fig. 4d, with amplitude indicated by grayscale and phase by color hue. Figure 4e synthesizes an all-in-focus image of the coin by stitching together the in-focus segments of the digitally refocused wavefields, demonstrating the system's capability to autonomously achieve focus across various depths[48]. Further exploitation of the refocused wavefields allows for the identification of regions with maximal contrast[44, 48], enabling the sample's 3D topography to be recovered. Displayed in Fig. 4f is the coin's reconstructed 3D map, elucidating the elevated features of the coin's surface, particularly the raised strips on the digit '1'. In Supplementary Video S4 and Fig. S12, we further show the 3D reconstruction process of the coin, showcasing SAPE's capability for imaging specimens with intricate 3D structures. These results highlight the potential of SAPE to be extended to distal-chip inspection endoscopes, offering a new tool for quality control and failure analysis in different manufacturing sectors.

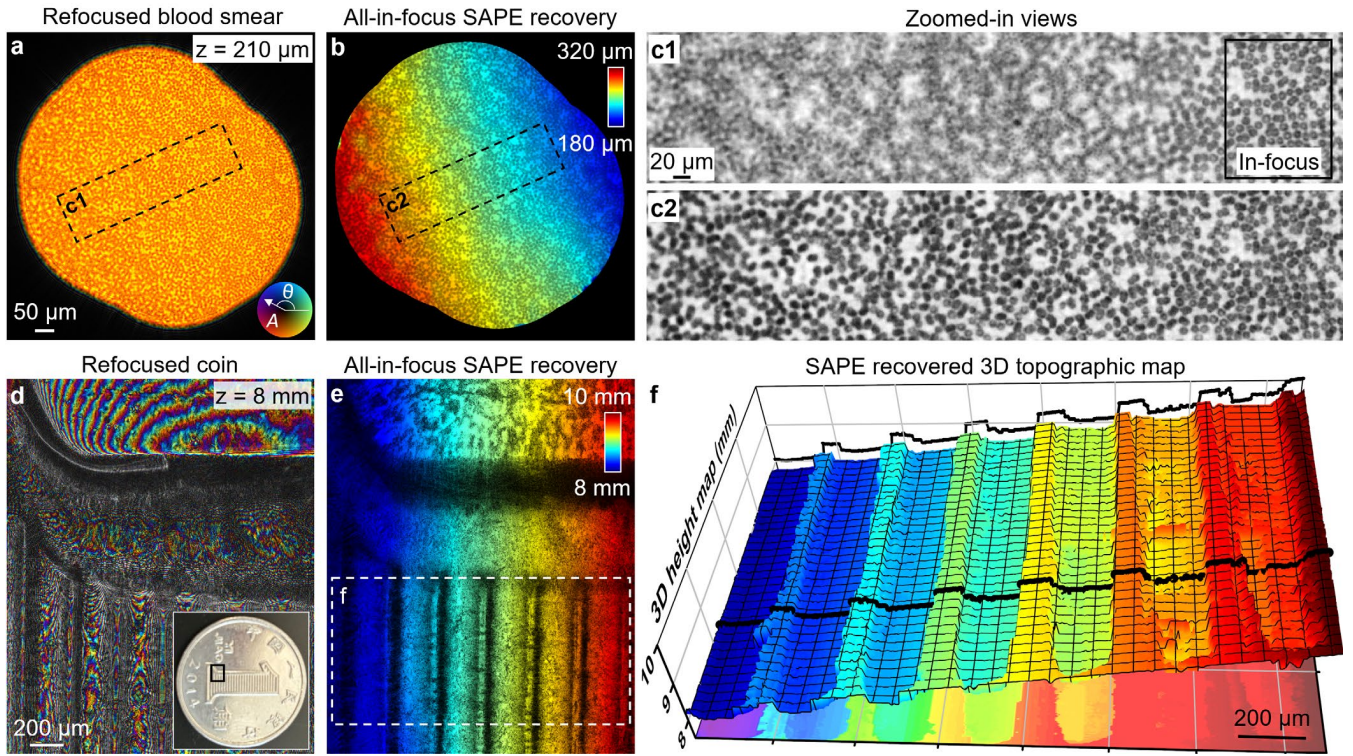

**Fig. 4| Proof-of-concept 3D topographic imaging demonstrations using SAPE.** (a) The recovered complex wavefield of a tilted blood smear slide, refocused at a depth of 210 μm, with amplitude indicated by grayscale and phase by color hue. (b) All-in-focus image of the blood smear slide, where the color map indicates the axial depth from 180 to 320 μm. (c) Zoomed-in views of (a) and (b). (d) The recovered complex wavefield of a coin. (e) All-in-focus image of the coin, where the color map indicates the axial depth. (f) The recovered 3D topographic map of the coin, which is obtained by locating the highest contrast of different refocused wavefields along the axial direction. Supplementary Video S3 shows the refocusing process of the tilted blood smear slide. Supplementary Video S4 shows the reconstruction process of the 3D surface profile of the coin.

## SAPE imaging beyond the diffraction limit

In conventional imaging systems, resolution is fundamentally restricted by the diffraction limit, defined by the imaging probe's maximum collection angle $\theta$. For a coherent endoscopic probe of radius $r$ and distance $d$ from the specimen, angle $\theta$ is determined by $\theta = \tan^{-1}(r/d)$, yielding a numerical aperture (NA) of $\sin(\tan^{-1}(r/d))$. The smallest resolvable linewidth is thus limited to $l = \lambda/(2 \cdot NA)$, where $\lambda$ is the illumination wavelength[49]. Unlike conventional endoscopy, SAPE leverages the handheld motion-induced shifts not only to acquire a ptychogram but also to construct a synthetic aperture that surpasses the physical constraints of the distal probe. Consequently, SAPE's achievable resolution is no longer limited by the maximum collection angle of the imaging probe. Instead, it is determined by the translational shift of the lensless distal facet. The synthetic NA in SAPE is a function of the translational area's radius ($R$) and the distance $d$ between the endoscope distal facet and the specimen: $NA_{syn} = \sin(tan^{-1}(R/d))$, resulting in a resolvable linewidth of $\lambda/(2 \cdot NA_{syn})$. We also note that when the specimen is placed close to the fiber bundle tip, the resolution is limited by the maximum collection angle of the fiber cores. For the employed fiber bundle, the measured NA of the fiber cores is approximately 0.42. It is possible to further improve this NA by attaching an additional coded surface on top of the fiber bundle tip. For implementing SAPE in distal-chip endoscopy, a properly designed coded surface can achieve a resolution equivalent to that of NA of 0.8[42], the highest in lensless ptychographic imaging with a theoretical limit of NA = 1.

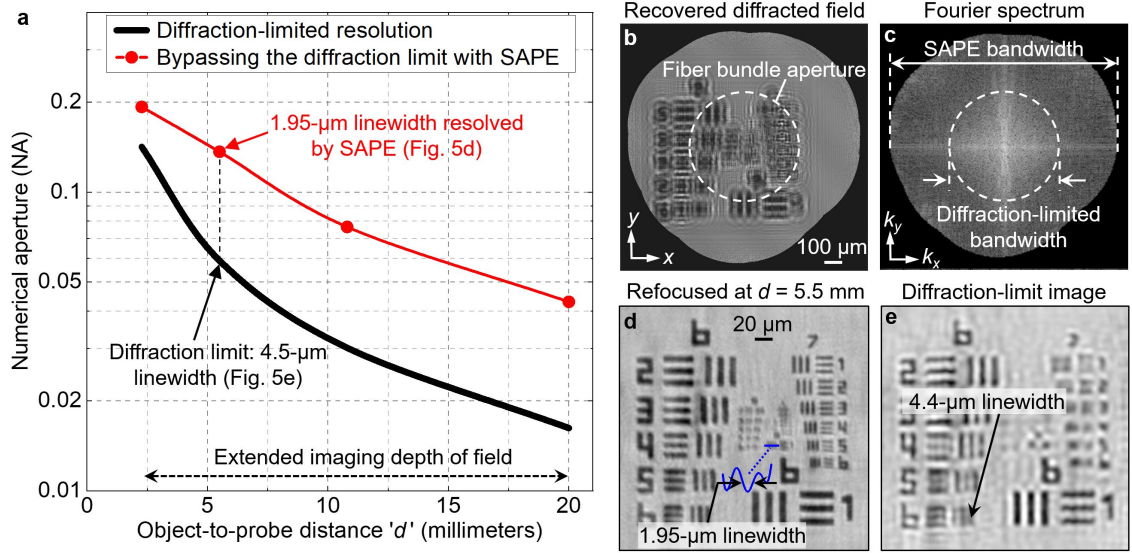

**Fig. 5| Surpassing the diffraction limit with SAPE.** (a) The black curve plots the diffraction-limited NA of conventional imaging systems, constrained by the maximum collection angle of the imaging probe. SAPE overcomes this limitation through the handheld motion of the fiber bundle tip, analogous to the aircraft movement in SAR. The translation motion not only facilitates the acquisition of a ptychogram but also synthesizes an imaging aperture exceeding the bundle's physical size. (b) The recovered diffracted wavefield $W(x, y)$ at the plane of the distal fiber bundle tip, located ~5.5 mm from the specimen. The dashed white circle represents the fiber bundle's physical aperture, while the wavefield synthesis at this plane facilitates aperture synthesis that exceeds the diffraction limit. (c) The Fourier spectrum of the wavefield $W(x, y)$ from (b), contrasting the SAPE's expanded bandwidth against the inherent diffraction-limited bandwidth. (d) SAPE reconstruction by propagating the recovered wavefield $W(x, y)$ for 5.5 mm, successfully resolving the 1.95-μm linewidth. The width of the entire group 8, element 1 is 10 μm. (e) A SAPE reconstruction with a minimum 1.5-μm scanning step size, approximating the diffraction-limited performance for comparison.

In Fig. 5, we performed an experiment to demonstrate SAPE's ability to surpass the diffraction limit across an extended depth of field. Figure 5a features a black curve that plots the diffraction-limited NA of the imaging probe against the object-to-probe distance $d$. Four SAPE reconstructions were performed at different object-to-probe distances, as denoted by the red dots in Fig. 5a. Remarkably, the synthetic NAs achieved in these reconstructions consistently exceed their diffraction-limited benchmarks. In Fig. 5b, we present the recovered diffracted wavefield $W(x, y)$ at the plane of the distal fiber bundle tip, with $d$ = 5.5 mm. The fiber bundle aperture in this figure is highlighted by the dashed white circle with a radius of 325 μm. The Fourier spectrum of the SAPE reconstruction is shown in Fig. 5c, where we compare SAPE's bandwidth with the inherent diffraction-limited bandwidth of the imaging probe. In the zoomed-in view of Fig. 5d, we digitally propagate the recovered wavefield to the object plane, successfully resolving a linewidth of 1.95 μm of group 8, element 1. For comparison, Fig. 5e shows a reconstruction that approximates the diffraction-limited performance of the employed imaging probe. In this experiment, the fiber bundle tip was translated using a motorized stage with a scanning step size of 1.5 μm. This step size results in a synthetic aperture that deviates only by ~3% from the original aperture size. The smallest resolvable linewidth in this case is 4.4 μm in Fig. 5e, closely matching the fiber bundle probe's diffraction-limited resolution of 4.5 μm. In Supplementary Fig. S13, we further compare the SAPE reconstructions with diffraction-limited performances at object-to-probe distances $d$ = 2.3 mm, 10.8 mm, and 20 mm. These comparisons not only validate SAPE's ability to surpass the diffraction limit of the employed imaging probe but also showcase its extended depth of field exceeding 2 cm. We also note that the area in the spatial domain (Fig. 5b) represents the synthetic aperture determined by the lateral coverage of the fiber bundle tip, while the area in the Fourier domain (Fig. 5c) represents the corresponding spatial-frequency coverage based on the wavevector components. In Supplementary Fig. S14, we perform an analysis to compare coverages in these two different domains.

## Discussion

SAPE integrates the concepts of SAR, lensless coded ptychography, and ultrathin endoscopy for micro-endoscopic imaging beyond the diffraction limit. Unlike other lens-based or lensless systems, SAPE captures reference-free coherent diffraction patterns from specimens using a handheld fiber bundle tip or a handheld distal-chip configuration. The densely-packed fiber cores at the bundle tip or the coded surface in the distal-chip design act as a structured ptychographic probe 'beam', modulating both the amplitude and phase of the diffracted wavefield over a confined area. In our implementation, the dynamic wavefront distortions caused by bending and hand motion are mitigated through low-rank spatiotemporal decomposition of the bundle's modulation profile. By using a fiber bundle with a 3.3-μm core-to-core distance, we demonstrated SAPE's ability to resolve features down to a 548-nm linewidth on a resolution target and to achieve centimeter-scale depth of field, facilitating the imaging of samples with complex, non-planar topographies. We have also validated SAPE's versatility by imaging reflective samples and demonstrated its capability of rendering 3D topographic maps.

Compared to conventional ptychographic implementations, SAPE offers several unique advantages. First, SAPE eliminates the need to model the interaction between the illumination beam and the sample, enabling the recovery of the exit wavefront that can be propagated to any axial position post-measurement. This allows for the imaging of a variety of samples, including those with non-planar surfaces and objects at various distances from the probe. In contrast, conventional ptychographic approaches often rely on a consistent illumination probe beam, limiting their applicability to flat samples at a fixed distance. Second, SAPE bypasses the resolution limit imposed by the core-to-core distance of the fiber bundle probe, achieving super-resolution imaging beyond the diffraction limit without the need for interferometric measurements. Our experiments demonstrate the ability to resolve a linewidth of 548 nm, which is among the finest resolutions reported in lensless endoscopic demonstrations. These unique features of SAPE expand its applicability to a wide range of practical samples and imaging scenarios, setting it apart from conventional ptychographic implementations. Moreover, the principles underlying SAPE have the potential to revolutionize the field of synthetic aperture imaging, paving the way for next-generation systems operating in the visible light spectrum.

In conventional endoscopic procedures, the distal end of the imaging probe is typically translated across the specimen to acquire multiple images, which are then stitched together to visualize a larger field of view. SAPE capitalizes on this routine operation but takes it a step further. The handheld motion of the distal end naturally translates to different lateral positions, allowing SAPE to not only broaden the field of view but also overcome the diffraction limit, address the issue of pixelation, and mitigate other optical constraints inherent to the lensless imaging probes. Furthermore, the elimination of interferometric measurements in SAPE simplifies the imaging process and enhances its adaptability to various environments.

An interesting question for SAPE is its applicability to more complex optical systems beyond the fiber bundle and coded surface discussed in this work. These complex systems are often characterized by transmission matrices (TMs)[50], which can be regarded as space-variant point spread functions of complex optical systems. In recent years, the ability to characterize TMs has driven emergent fields in optics such as imaging through scattering media and multimode fibres[51, 52, 53, 54]. Our future efforts will focus on the extension of SAPE to the characterization of multimode fiber. This involves replacing the multiplicative assumption governing the interaction of the fiber bundle or coded surface with the diffracted wavefield with a more general description involving TMs. We envision applications in the characterization of long-range multimode interconnects like optical communication links, where the use of an external holographic reference wave, as commonly used in the characterization of TMs, is inhibited by extraneous instabilities such as turbulence.

The SAPE method has proven effective yet has room for improvement. As robotic endoscopic surgery gains popularity and becomes increasingly sophisticated, the integration of SAPE with robotic systems presents an exciting frontier. Employing a robotic system could offer precision control in the translation of the probe, potentially enhancing the quality of the acquired images and further pushing the boundaries of resolution. Another future direction is to explore SAPE's potential in a side-view configuration. By rotating a 45-degree prism on the lensless endoscope probe, SAPE can acquire 360-degree views of the surrounding objects, similar to certain endoscopic optical coherence tomography configurations[55]. In our existing setup, the reconstruction algorithm does not account for small hand motion along the axial direction, which can lead to image degradation when compared to results obtained with a motorized stage. However, the rich information provided by the ptychogram offers a pathway for improvement via algorithmic correction of minor defocusing shifts during the acquisition process. As demonstrated in Supplementary Fig. S18, the defocusing effect can be partially compensated by the spatiotemporal decomposition of the endoscope tip's modulation profile. Supplementary Fig. S19 further quantifies the impact of axial hand movements on the reconstruction quality. The results reveal that SAPE is generally robust against small axial movements, maintaining high reconstruction quality. These findings guide future research efforts in developing more robust and motion-tolerant SAPE systems. For example, based on the decomposed orthogonal modes, it may be possible to track the axial positions of the imaging probe post-measurement. Lastly, the principles underlying SAPE have broader implications, including potential enhancements to SAR, Fourier ptychographic microscopy and tomography[56, 57, 58], and far-field synthetic aperture imaging[46, 59, 60]. For instance, by adopting SAPE's approach of synthesizing a larger virtual aperture through movement, SAR could extend its operational capabilities to optical wavelengths[61], opening new avenues for remote sensing applications with autonomous vehicles and drones.

## Materials and methods
### Experimental setup
In our implementation with lensless fiber bundle probe (Fujikua, FIGH-30-650S), we place the distal end of the bundle close to the specimen to acquire the diffraction data. The proximal end of the fiber bundle is then imaged onto a camera using a custom-built microscope system, which includes a 10×, 0.4-NA Nikon objective lens and a tube lens. The employed camera acquires images at a rate of 100 frames per second. An important consideration of the SAPE setup is the defocusing of the camera by a small distance $h$. This defocusing distance is essential for converting the complex wavefields from the proximal end of the fiber bundle into intensity variations for detection, as discussed in Supplementary Note 2. A small defocus value is insufficient for converting the phase information into intensity variations. Conversely, a large defocus value introduces diffraction effects that complicate the tracking of object shifts. The chosen defocus distance, set at ~20 μm in our experiments, balances these two considerations. SAPE reconstructions with different defocus distances are also presented in Supplementary Fig. S3. Alternatively, SAPE can be implemented in an existing distal-chip endoscope with a miniaturized camera. The small lens element of the miniaturized camera is then replaced with a coded surface[18] that serves the role of the fiber bundle for light wave modulation. In this case, the intentional defocusing distance corresponds to the space between the coded surface and the pixel array of the image sensor. The acquisition time of SAPE varies depending on the total number of acquisitions, ranging from approximately 0.5 to 5 seconds, with the number of raw measurements spanning from 50 to 500. SAPE reconstructions using different numbers of acquisitions are detailed in Supplementary Figs. S10-S11, illustrating the effects of these variables on image quality and reconstruction.

### Positional tracking of the handheld endoscope probe
In SAPE, accurately tracking the positional shifts of the lensless distal tip is essential for reconstructing the object's diffracted wavefield $W(x, y)$ entering the imaging probe. As outlined in Supplementary Note 1, we begin with an initial cross-correlation analysis to distinguish between the correlation peaks from the correlation of the modulation profiles at different images and those resulting from the distal probe's shifts. A multi-reference strategy is then implemented to effectively capture positional shifts beyond the initial imaging area, a key step for synthesizing an aperture larger than the physical dimension of the lensless distal tip. This process culminates in refining the reference images by summing the shifted versions of the captured images[38]. With adequate training, an operator can effectively control the range of hand motion to optimize the experiment's outcome. Supplementary Fig. S15 illustrates two different trajectories of the handheld probe during operation. Supplementary Fig. S16a shows the comparison between the positions recorded by the motorized stage (illustrated as green dots) and the corresponding estimated positions (illustrated as red dots). The analysis reveals an average discrepancy of approximately 0.27 microns between the estimated positions and the ground truth, as shown in Supplementary Fig. S16b. This small discrepancy indicates a high degree of precision in our estimation process, despite the challenges posed by the fiber core pixelation.

**SAPE reconstruction**
In SAPE, it is beneficial to measure the modulation profile of the fiber bundle or the distal coded sensor beforehand to serve as the initial profile for the reconstruction process. This calibration experiment can be performed using an object with rich spatial features, such as a blood smear, to recover the modulation profile of the lensless imaging probe. Compared to the real experiment with an unknown specimen, the acquisition time in this calibration experiment is not an issue, allowing us to capture more images for better joint reconstruction of both the target and the modulation profile. Furthermore, in the calibration experiment, we can use mechanical stages to translate the calibration target with better precision and control, as opposed to relying on hand motions. Once the initial modulation profile is obtained, low-rank decomposition plays an important role in handling the variations in the modulation profile caused by hand motion and other system perturbations during the actual data acquisition process. By assuming that these changes are gradual and minor over time, we can effectively capture and represent the variations using a compact set of basis modes derived from the low-rank decomposition. Specifically, we apply singular value decomposition to the time-varying modulation profiles, projecting them into a lower-dimensional space. This process identifies a set of orthogonal modes that capture the essential variations in the modulation profile over time. The low-rank decomposition leverages the inherent redundancy in the modulation profile's variations, reducing the complexity of directly capturing and modeling these variations during the actual experiment. This simplification enables SAPE to handle the dynamic nature of the modulation profile of the lensless endoscope probe.

As detailed in Supplementary Note 3, the reconstruction procedure begins with the initialization of the object's diffracted wavefield $W(x, y)$ and the modulation profile of the lensless imaging probe. The subsequent steps involve cropping sub-regions from the wavefield $W(x, y)$ and conducting point-wise multiplication with the modulation profile of the bundle or the coded surface. The resulting modulated exit waves are then propagated to the sensor plane and updated by the captured intensity measurements. Subsequently, these updated wavefields are back-propagated to the plane of the distal facet, where both the diffracted wavefield $W(x, y)$ and the modulation profile of the endoscope facet are refined using the rPIE algorithm[62]. The final step projects the reconstructed position-dependent modulation profiles into a lower-dimensional space using singular value decomposition. Supplementary Fig. S17 shows the orthogonal modes of the fiber bundle profile and its evolution during the image acquisition process. A simulation study in Supplementary Fig. S18 shows that the slight defocusing effect can be modeled and corrected by the modulation profiles of the endoscope tip. Another consideration for SAPE reconstruction is the signal-to-noise ratio (SNR). Different from fluorescence imaging in photon-limited settings, the SNR here is determined by the illumination power and the full well capacity of the image sensor. In our experiments, we typically adjust the laser power to barely saturate the 8-bit camera, ensuring that the maximum intensity of the raw image reaches 255. As shown in Supplementary Fig. S20, the maximum SNR achieved is approximately 14, which depends on the signal strength of the input intensity values.

**3D topographic reconstruction**
In the process of 3D topographic reconstruction, we start by digitally propagating the recovered diffracted wavefield from the lensless endoscope probe toward the object plane, generating z-stack images at different axial positions. This image stack enables us to pinpoint the axial in-focus positions where phase contrast is maximized[44, 48]. As demonstrated in Supplementary Fig. S12 and Video S4, it can generate a detailed 3D map for each pixel within the image, highlighting topographical features such as the raised elements and intricate surface details.

**Mouse colon sample preparation**
We used C57BL/6 mice aged 8-12 weeks and weighing 20-30 g, sourced from Taconic, Germantown, NJ. These mice were first anesthetized with isoflurane inhalation and then humanely euthanized via exsanguination by perforating the right atrium. Following euthanasia, transcardial perfusion was performed using ice-cold, oxygenated Krebs solution (95% $O_2$, 5% $CO_2$), which comprised 117.9 mM NaCl, 4.7 mM KCl, 25 mM $NaHCO_3$, 1.3 mM $NaH_2PO_4$, 1.2 mM $MgSO_4$, 2.5 mM $CaCl_2$, and 11.1 mM D-glucose. The distal colon was then excised, thoroughly cleansed with the same Krebs solution to remove fecal matter, and then longitudinally sliced and rolled into a swiss roll configuration. This prepared tissue was then mounted on a metal substrate holder and fixed in 4% paraformaldehyde within a 0.16 M phosphate buffer solution containing 14% picric acid (Sigma). All experimental procedures involving animals were rigorously reviewed and received approval from the University of Connecticut Institutional Animal Care and Use Committee.

**Acknowledgments**
This work was partially supported by the National Institute of Health R01-EB034744 (G. Z.), the UConn SPARK grant (G. Z.), National Science Foundation 2012140 (G. Z.), and the National Institute of Health U01-NS113873 (B. F. and G. Z.). P. S. acknowledges the support of the Thermo Fisher Scientific fellowship.

**Conflict of interest**
The authors declare no competing interests.

**Author contributions**
G. Z. conceived the concept of SAPE. P. S. and R. W. developed the prototype systems, conducted the experiments, and analyzed the data. P. S., R. W., and Y. L. prepared the display items. J. L. and B. F. prepared the mouse colon samples. All authors contributed to the writing and revision of the manuscript.

**Data availability**
All the data and methods needed to evaluate the conclusions of this work are presented in the main text and Supplementary Materials. Additional data can be requested from the corresponding authors.

**Supplementary information**
Supplementary information accompanies the manuscript can be found on the *Light: Science & Applications* website (http://www.nature.com/lsa)